# Structural evidence of magnetic field induced devitrification of kinetically arrested antiferromagnetic phase in $La_{0.175}Pr_{0.45}Ca_{0.375}MnO_3$: A Low-temperature high-magnetic field x-ray diffraction study

Shivani Sharma, Aga Shahee, Kiran Singh, N. P. Lalla[*] and P. Chaddah

UGC-DAE Consortium for Scientific research, Indore-452001

## Abstract

The low-temperature and high-magnetic field (2K, 8T) powder x-ray diffraction (LTHM-XRD) measurements have been carried out at different temperatures (T) and magnetic fields (H) to investigate the structural phase diagram for phase separated $La_{0.175}Pr_{0.45}Ca_{0.375}MnO_3$ (LPCMO) manganite. The antiferromagnetic (AFM) $P2_1/m$ insulating phase undergoes field induced transformation to ferromagnetic (FM) *Pnma* metallic ground state below its AFM ordering temperature (220K) in zero-field cooling (ZFC) from room temperature. At T≤25K, the field induced FM *Pnma* phase remained irreversible even after complete removal of field. However, at higher T (~39-65K), the field induced transformation is partially reversible. This behaviour has been attributed to magnetic field induced devitrification of the glass-like arrested AFM $P2_1/m$ phase to FM *Pnma* equilibrium phase. The devitrified FM *Pnma* phase starts transforming back to AFM $P2_1/m$ phase around ~39K on heating the sample under zero field. Our results corroborate the evidence of strong magneto-structural coupling in this system. An H-T phase-diagram has been constructed based on LTHM-XTD data, which resembles with the one made from magnetic measurements. These results have been explained on the basis of kinetic arrest of first order phase transition and field induced devitrification of the arrested phase.

**Introduction:**

The quest of understanding the physics of glasses in condensed matter has led researchers to conceptualize unconventional glasses like "magnetic-glasses"[1,2,3,4,5] and "strain glasses" [6,7,8]. While the basic concept still pivots around the "critical slowing down" of the transformation kinetics, like in conventional liquid to crystal first-order phase transition, the idea of glass-like kinetic-arrest has been extended even in the case of structurally order-order phase transitions[9]. In this regard, rigorous efforts have been made to understand the kinetics of such arrested phase using unconventional measurement protocols in (magnetic field-temperature) H-T space for different materials[1,2,3,4,5,10,11,12]. In manganites, the understanding of the phase coexistence and phase separation (PS) is very important to affirm the microscopic origin of colossal magnetoresistance (CMR) behaviour. Out of various manganites, the charge ordered (CO) $La_{5/8-y}Pr_yCa_{3/8}MnO_3$ (LPCMO) has gained special attention particularly to understand its PS and frozen state behaviour[7,8,13,14,15,16,17]. Different techniques along with magnetic field (H) as an additional perturbation with temperature (T), has been used to understand the frozen/glass-like behaviour in LPCMO manganite like magnetization[8,13,18], electrical resistivity[15,16], magnetic force microscopy(MFM)[7,16], photoemission electron microscopy (PEEM), resonant elastic soft x-ray scattering (RSXS)[19], nuclear magnetic resonance (NMR)[20] etc. It is reported that the x-ray and electron beam irradiation can also de-arrest the arrested phase in manganites[9,13].

The metal insulator transition temperature in LPCMO is very sensitive to elemental composition. Even a slight variation in composition can change the transition temperature effectively[15]. In case of LPCMO, many reports present contradictory results on phase separation scenario. Kiryukhin et al.[13] and Lee et al.[21] suggested the multiphase coexistence in low temperature regime. Later on Muńoz et al[22] have discarded the possibility of three different segregated phases using neutron diffraction and synchrotron based XRD. They have also mentioned that the differences in the dimensions of CO-AFM and FM cells are very small and difficult to resolve. The detailed microscopic studies of magnetic PS scenario have been done by Wu et al.[7], Zhang et al.[14] and Rawat et al.[16] using MFM, magnetization and transport measurements on slightly varying compositions to explore the kinetically driven glass transition in LPCMO. According to Wu et al.[7], the glass state is tightly associated with the supercooled state of the first-order CO AFM to FM phase transition, where the kinetics of the phase transition is strongly influenced by the accommodation strain.

The kinetically arrested glassy state in LPCMO has been explored rigorously through transport and microscopic studies like MFM. The de-arrest of kinetically arrested high temperature (HT) AFM state starts at different values of H for different temperature and also depends on the path followed to reach the particular temperature. In all these studies, the main focus was to map the FM and AFM regions which results due to different kind of spin interactions but these studies did not provide any information about the crystal lattice. The AFM and FM states have different crystal structures and one can see the arrest and de-arrest of *crystal structure* of AFM/FM phases by performing XRD/neutron at different temperature and magnetic field. Recently, such experiments have been performed on $Ni_{37}Co_{11}Mn_{42.5}Sn_{9.5}$ and $La_{0.5}Ca_{0.5}MnO_3$ by using in-field neutron diffraction studies[23]. In spite of various microscopic evidence of PS till date, there is no structural evidence of arrest and de-arrest of HT AFM phase in LPCMO. The true structural nature of this system in the presence of magnetic field is yet to be unravelled. Keeping all this information in view, in the present work, we have carried out detailed investigation on LPCMO using LTHM-XRD to illustrate the field induced devitrification of kinetically arrested AFM ($P2_1/m$) phase. This work will highlight the in-field structural aspects of low temperature (LT) frozen/glass state in LPCMO as pointed out by Wu et. al[7] and Rawat et al.[16]. We have also constructed the H-T phase diagram for crystal structure of $La_{0.175}Pr_{0.45}Ca_{0.375}MnO_3$ based on LTHM-XRD results which is well in agreement with the existing literature[7,20,22].

**Experimental:**

Polycrystalline sample of $La_{0.175}Pr_{0.45}Ca_{0.375}MnO_3$ (LPCMO) has been prepared via solid state reaction route. The starting ingredients $La_2O_3$, $Pr_6O_{11}$, $CaCO_3$ and $MnO_2$ were preheated before weighing. The preheated ingredients were mixed in stoichiometry ratio using agate mortar pastel. The mixture was calcined at 1000°C for 24 hrs. The powder was then pressed into the pellets of 10mm diameter and final sintering was done at 1300°C. Room temperature (RT) structural characterization has been done using Rigaku's XRD with Cu-K$_\alpha$ radiation. Oxygen stoichiometry was determined using iodometric titration. The occurrence of charge ordering in the sample was investigated using simple low-temperature XRD setup. In-field x-ray diffraction studies have been performed using the recently developed low temperature and high magnetic field XRD (LTHM-XRD) setup operated at 12kW. The details of this setup are given in Ref.[24]. For performing magnetic field dependent XRD measurements with field cycling between 0-8T at various fixed temperatures, each time the sample is zero field cooled (ZFC) from RT to the desired temperature to avoid any possible history dependent effect. After reaching the desired T, the field is raised to the desired values and the XRD data is recorded. The same procedure is followed in decreasing field cycle also. After the completion of 0-8T-0 field cycle, the XRD was then recorded during warming under zero field to investigate the possible melting of low temperature FM phase.

**Results and Discussions:**

Figure 1a shows the Rietveld refined RT XRD pattern of LPCMO fitted with *Pnma* space group. The absence of any unaccounted peak confirms the single phase nature of the as prepared LPCMO sample. The iodometric titration confirm that the as prepared $La_{0.175}Pr_{0.45}Ca_{0.375}MnO_{3-\delta}$ sample is nearly stoichiometric with $\delta=0.02\pm0.01$, which is well within the titration error. With decreasing T in ZFC condition, the sample undergoes paramagnetic (PM) (*Pnma*) to CO AFM (*P2$_1$/m*) phase transition at 220K , which persist down to 2K in ZFC condition. The occurrence of PM (*Pnma*) to CO AFM (*P2$_1$/m*) structural transition below $T_{CO}$ = 220K is typified by the appearance of CO superlattice peaks as can be seen clearly in temperature dependent XRD in the inset (i) of figure 1b. The corresponding change in the perovskite lattice can be seen in inset (ii) of figure 1b. As reported by Muńoz et al.[22] we could not detect any phase coexistence within our resolution limit. They have observed small but finite FM phase fraction coexisting with the major AFM phase below $T_{CO}$ = 220K for y=0.35, which increases below insulator to metallic transition down to 15K. This appearance of FM fraction below insulator to metallic transition decreases with increasing Pr content (x=0.375). In our case, the Pr content is 0.45 which is higher than the composition of the phase co-existence regime as reported in Ref.22. We believe that there is no FM phase fraction below 220K in ZFC condition.

At 2K (ZFC), we have investigated the change in structural properties as a function of magnetic field. With the application of field, one can see a clear effect on diffraction peak height and peak positions, as shown in figure 2a. These results show first order CO AFM (*P2$_1$/m*) to FM (*Pnma)* phase transition. During increasing field cycle, the AFM phase starts transforming to FM phase at 2T, and the transformation is almost complete at 5T, as there is no change in the peak intensity of FM phase with further increasing H. While decreasing field cycle, interestingly, the field induced FM phase did not revert back to CO AFM phase and remains FM, see figure 2b. After the completion of field cycle (increasing-decreasing, 0-8T-0) the sample was heated above its fixed temperature to see the possible structural changes. Such measurements were performed at various temperatures e.g. 2K, 8K, 15K, 25K, 39K,

50K, 80K, 106K, 125K, 150K and 192K. Few of these results are presented in figure 2. Figure 2(a-f) illustrates that below 39K, the field induced FM phase did not revert back to AFM state i.e. this transformation shows irreversible behaviour, whereas at 39K≤T≤65K, this field induced transformation is partially reversible and above that, it is completely reversible. It was observed that the critical field ($H_c$) required to transform AFM phase to FM phase is strongly temperature dependent. The field induced transition is rather broad in H, i.e. $H_c$ has two values namely $H_{c1}$ (the field at which AFM *P2$_1$/m*-FM *Pnma* phase transition starts) and $H_{c2}$ (the field at which the phase transition completes), showing phase coexistence in between them. Figure 3 shows the H-T phase diagram constructed based on LTHM-XRD measurements, showing the variation of $H_{c1}$ and $H_{c2}$ during increasing field. Between 2K and 192K, the $H_c$ first decreases with T and then increases, having minima at ~39K. The $H_{c1}$ and $H_{c2}$ limit the phase coexistence region. Similar $H_{c3}$ and $H_{c4}$ also exist during decreasing field cycle. However such values ($H_{c3}$ and $H_{c4}$) are observed only for T≥39K where the phase transformation exhibits hysteretic and reversible behaviour during isothermal field cycle.

The variation in the FM phase fraction during field cycling has been determined and presented at some selected temperatures below 39K; see figure 4(a-d). The structural phase change on heating after each field cycle is also shown in figure 5a,b. The FM *Pnma* phase transforms back to AFM *P2$_1$/m* phase on heating. Figure 5c exhibits the transformation behaviour and the corresponding FM phase fraction variation on heating after field cycling at 2K, 15K and 25K. From figure 5c, it can be seen that irrespective of the isothermal field cycling temperature (selected below 35K), the melting of FM *Pnma* phase starts at ~30K and completes at ~65K. The CO AFM phase remains stable at T ranging from 65-220K and above 220K, it further transforms to PM state through a first order structural phase transition.

The features of the observed H-T phase diagram as shown in figure 3 can be explained on the basis of kinetic arrest scenario. Figure 6 shows the proposed H-T phase diagram for disordered broadened first order phase transitions. Due to disorder broadening[25], the phase transition lines in the H-T plane will convert into bands. Following the notations of Kranti et al.[18] three bands namely supercooling (H*,T*), superheating (H**,T**) and kinetic-arrest (freezing) ($H_k$,$T_k$) bands have been assigned. The increase of the $H_{c1}$ with lowering T and the irreversible nature of the transformation of LT AFM phase into LT FM phase (below 39K) indicate that it is the LT FM phase which is ground state and not the LT AFM phase. Due to arrest of the transformation kinetics at ~65K during cooling, the AFM phase leads to kinetically frozen state down to the lowest T under ZFC condition (path 1). Only after the application of a certain field, the glassy AFM phase devitrifies to LT FM phase. The kinetic arrest of HT AFM phase and its devitrification to FM ground state on field application has been reported in $Pr_{0.5}Ca_{0.5}MnO_3$ by Banerjee et al.[26] through magnetic and magneto-transport studies. At low fields (0T- ~1T), the ($H_k$,$T_k$) band lie above the (H*,T*) band as the results AFM $P2_1/m$ phase in ZFC state remains completely arrested down to the lowest T[18]. The area within the (H*,T*) and (H**,T**) band represents the phase separated (PS) regions. For more explanation, let us consider another path 2′, here the HT-AFM phase starts transforming to LT-FM states ~1T (lower value end of H*,T* band) and transforms completely at ~3T (higher value end H*,T* band) but with decreasing H, it follows the path 3′. Along it, the LT-FM state starts transforming back to HT-AFM state around 1T (higher value end of H**,T** band) but this transition doesn't complete even down to 0T and we get a phase coexistence region here. However, at higher T, for example T>65K, the complete AFM to FM and FM to AFM transition takes place in forward (path 2″) and reverse (3″) directions of field cycles. These results thus confirm that at low temperature the FM phase is the ground state and AFM phase is the kinetically frozen state for this system. This phase diagram thus completely follows the kinetic arrest behaviour of FM ground state as explained in earlier reports[18,27].

In conclusion, our detailed structural investigation using LTHM-XRD shows the presence of strong magneto-structural coupling in LPCMO system. In the studied composition the HT equilibrium CO AFM phase gets arrested below 39K under ZFC condition. However, with the application of magnetic field devitrification of AFM $P2_1/m$ state takes place resulting into LT FM *Pnma* phase, which remains irreversible at T≤39K even after complete removal of the field. The irreversibility indicates that the transformed phase is a ground state phase. It is partially reversible at intermediate temperatures (~39-65K) and completely reversible above T>65K. Our results clearly demonstrate the arrest and de-arrest of *crystal structure* in LPCMO which is consistent with the earlier phase diagram, constructed on the basis of transport, magnetization and MFM results. Such studies will be very useful to know the actual ground state structure and field driven magneto-structural phase transition in other systems too. This also highlights the importance of the knowledge of crystal structure to understand the complex magnetic phase diagram of phase separated manganites.

**Acknowledgements:**

Authors gratefully thank Director Dr. A. K. Sinha, the Centre Director Dr. V. Ganesan and the former Centre Director Prof. A. Gupta for their constant support and encouragements to development of LTHM-XRD facility. Authors also like to thank Dr. A. Banerjee for useful discussion regarding magnetic shielding for LTHM-XRD facility. NPL and KS are very much thankful to Dr. R. Rawat, Er. P. Sarvanan and Ms. Preeti Bhardwaj for their frequent help during magnet cooling and also thankful to Mr. N. L. Ghodke for technical help regarding hardware/software for x-ray shutter control.

**Figure captions:**

**Figure 1:** (a) Reitveld refined XRD pattern of LPCMO at 300K and (b) temperature dependent XRD patterns of LPCMO down to 2K; the inset (i) shows the zoomed view of charge ordered peak and (ii) shows the transformation from PM (*Pnma*) to AFM (*P*2$_1$/*m*) state.

**Figure 2:** Magnetic field dependent XRD measurement during isothermal increasing and decreasing field cycling at 2K, 25K, 39K, 50K, 106K and 192K, each time after ZFC condition.

**Figure 3:** H-T phase-diagram constructed using the experimental XRD data during field increasing field cycle.

**Figure 4:** The variation of FM phase fraction with H at (a) 2K, (b) 39K, (c) 50K and (d) 192K. The irreversibility below 39K and transformation hysteresis above 39K is clearly evident here.

**Figure 5:** T- dependent XRD pattern of LPCMO after performing field cycling at (a) 2K and (b) 15K, (c) 25K and (d) the variation in the FM phase fraction during heating (after performing field cycling at 2, 15 and 25K).

**Figure 6:** H-T phase diagram of LPCMO constructed based on data of isothermal increasing-decreasing field cycling after zero-filed cooling from RT to the particular temperature. The letters LT and HT corresponds to low temperature FM (*Pnma*) ground state phase and high temperature AFM (*P*2$_1$/*m*) phase, respectively. ($H_k$,$T_k$), ($H^*$,$T^*$), ($H^{**}$,$T^{**}$) represents the kinetic-arrest, supercooling and superheating band respectively. The upwards and downwards arrows denote the direction of magnetic field variation, respectively and PS represents phase separation.

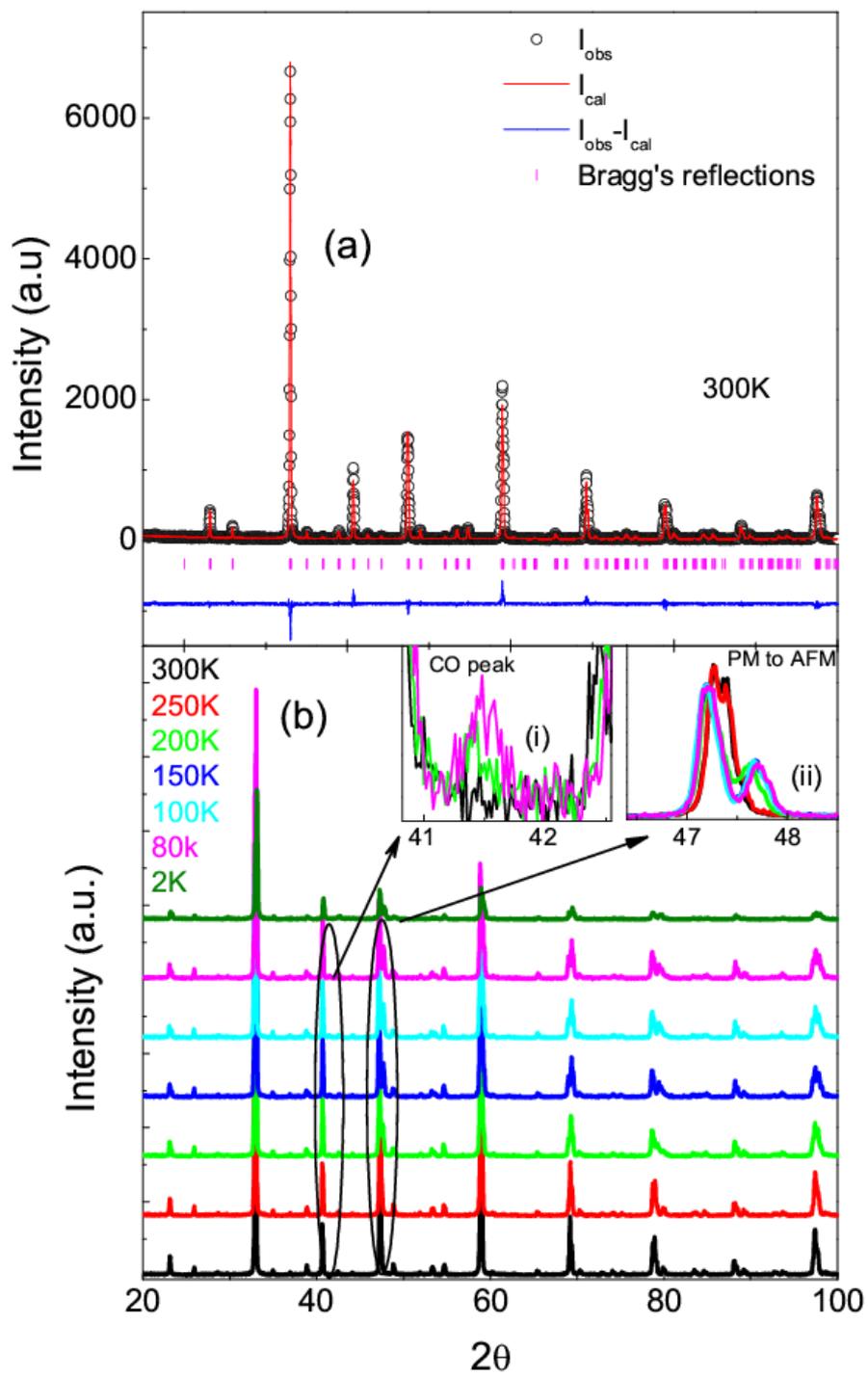

**Figure 1**

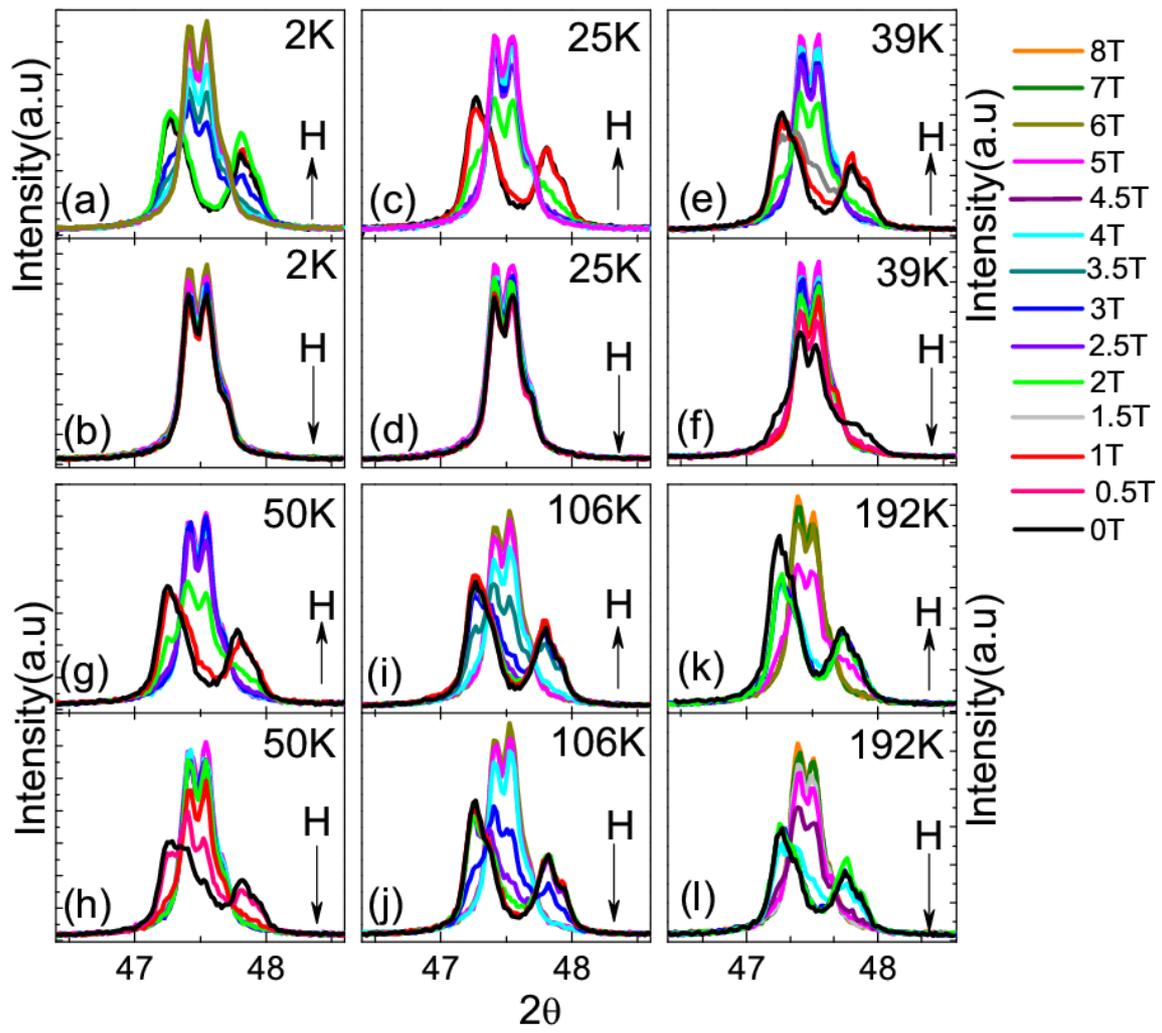

**Figure2**

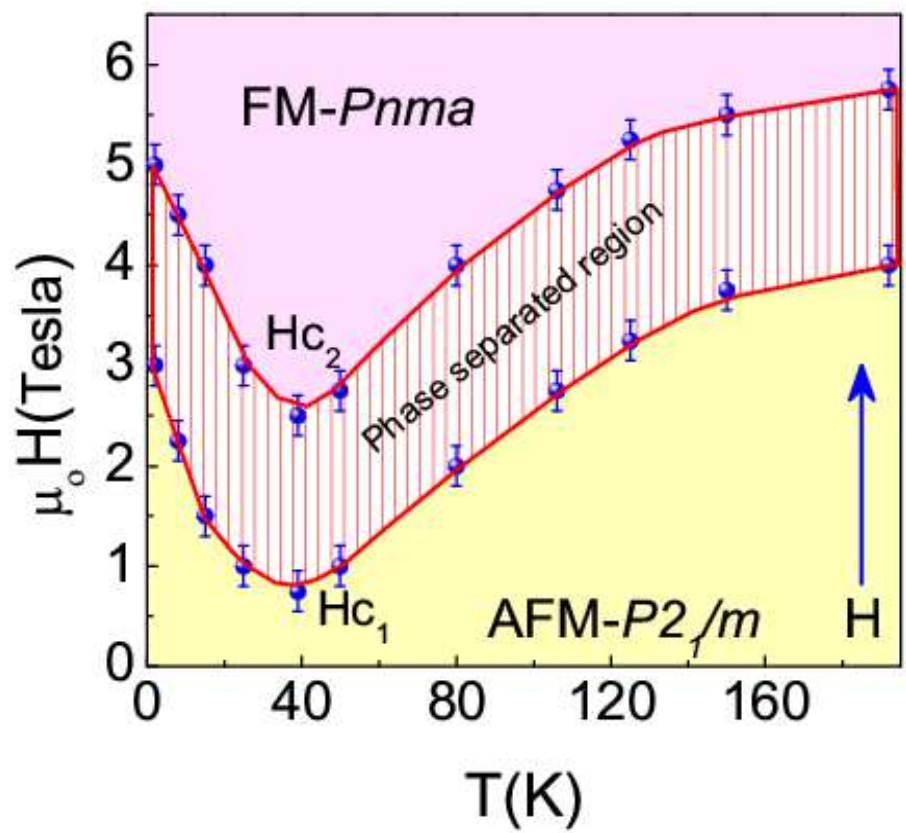

**Figure 3**

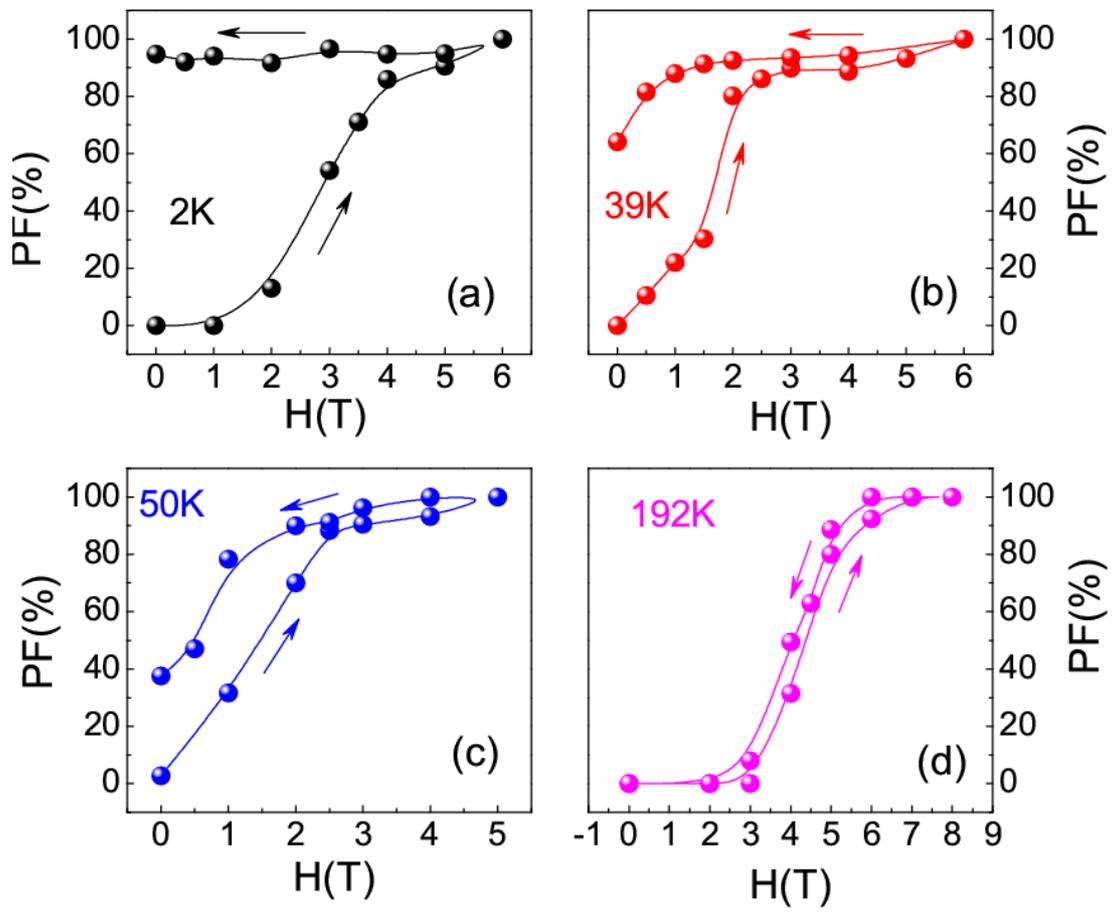

**Figure 4**

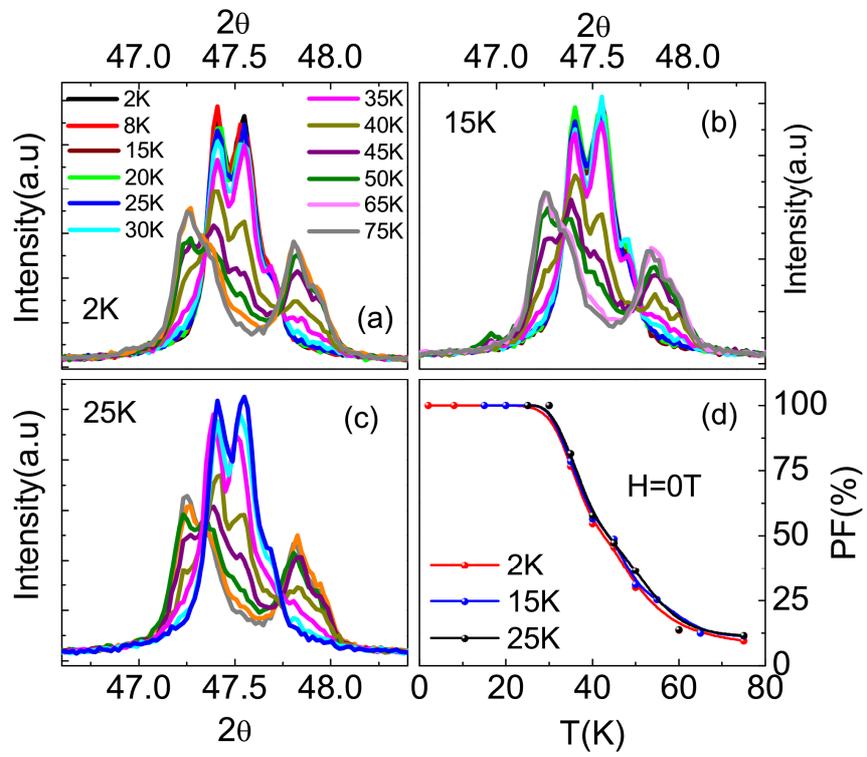

Figure 5

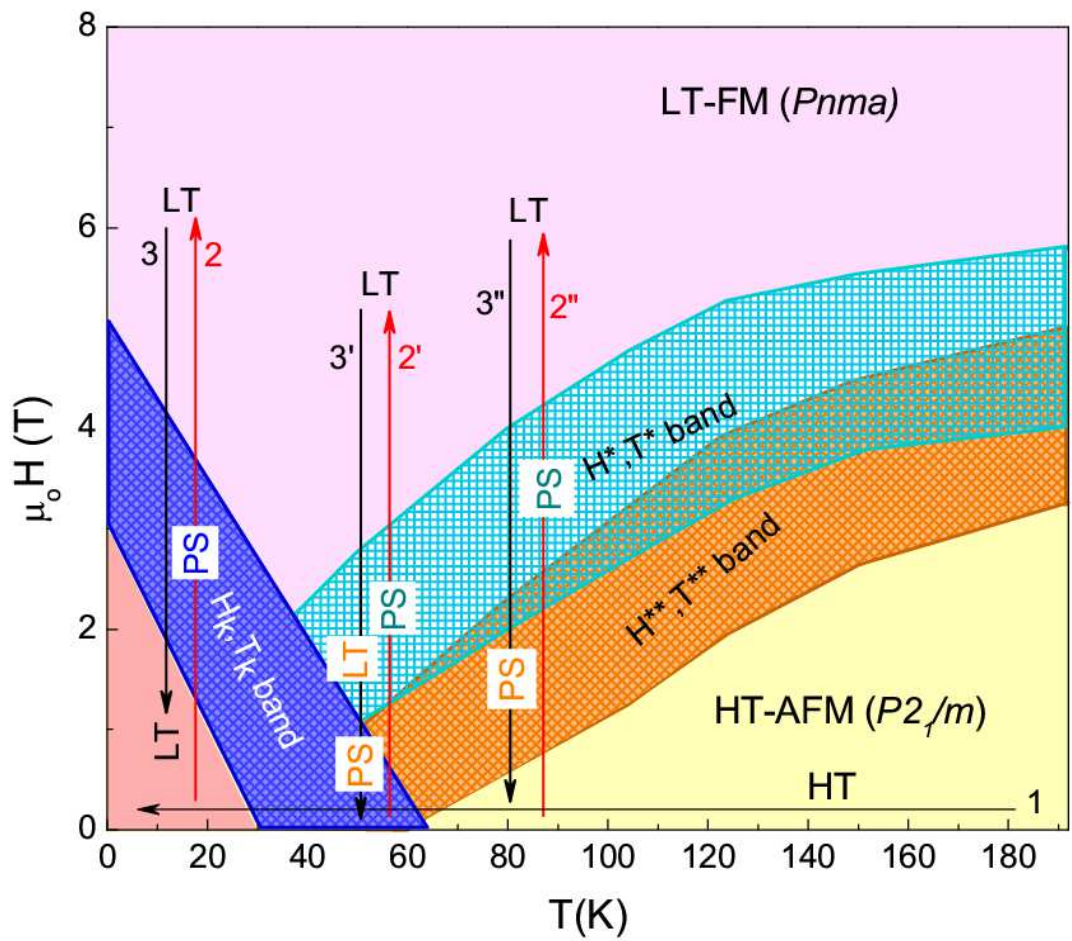

**Figure 6**

**References:**

1.  M. K. Chattopadhyay, S. B. Roy, and P. Chaddah, Phys. Rev. B **72**, 180401 (2005).

2. A. Banerjee, A. K. Pramanik, K. Kumar, and P. Chaddah, J. Phys.: Condens. Matter **18**, L605 (2006).

3. S. B. Roy, M. K. Chattopadhyay, P. Chaddah, J. D. Moore, G. K. Perkins, L. F. Cohen, K. A. Gschneidner Jr., and V. K. Pecharsky, Phys. Rev. B **74**, 012403 (2006).

4. S. B. Roy, M. K. Chattopadhyay, A. Banerjee, P. Chaddah, J. D. Moore, G. K. Perkins, L. F. Cohen, K. A. Gschneidner Jr., and V. K. Pecharsky, Phys. Rev. B **75**, 184410 ( 2007).

5. A. Banerjee, K. Kumar, and P. Chaddah, J. Phys.: Condens. Matter **20,** 255245(2008).

6. P. A. Sharma, S. B. Kim, T. Y. Koo, S. Guha, and S-W.Cheong, Phys. Rev. B **71**, 224416 (2005).

7. W. Wu, C. Israel, N. Hur, S. Park, S.-W.Cheong, and A. de Lozanne, Nat. Mater. **5**, 881 (2006).

8. L. Ghivelder, and F. Parisi, Phys. Rev. B **71**, 184425 (2005).

9. A. Shahee, D. Kumar, C. Shekhar, and N. P. Lalla, J Phys.: Condens. Matter **24**, 225405 (2012).

10. P. Chaddah, K. Kumar, and A. Banerjee, Phys. Rev. B **77**, 100402(R) (2008).

11. A. Banerjee, K. Mukherjee, K. Kumar, and P. Chaddah, Phys. Rev. B **74**, 224445 (2006).

12 . K. Sengupta, and E. V. Sampathkumaran, Phys. Rev. B **73**, 20406 (2006).

13. V. Kiryukhin, B. G. Kim, V. Podzorov, S-W. Cheong, T. Y. Koo, J. P. Hill, I. Moon, and Y. H. Jeong, Phys. Rev. B **63**, 024420 (2000).

14. L. Zhang, C. Israel, A. Biswas, R. L. Greene, and A. d. Lozanne, Science **298**, 805 (2002).

15. M. Uehara, S. Mori, C. H. Chen, and S.-W. Cheong, Nature **399**, 560 (1999).

16. R. Rawat, P. Kushwaha, D. K. Mishra, and V. G. Sathe, Phys. Rev. B **87**, 064412 (2013).

17. D. K. Mishra, V. G. Sathe, R. Rawat, and V. Ganesan, J. Phys.: Condens. Matter **25,** 175003 (2013).

18.  K. Kumar, A. K. Pramanik, A. Banerjee, P. Chaddah, S. B. Roy, S. Park, C. L. Zhang, and S.-W. Cheong, Phys. Rev. B **73**, 184435 (2006).

19. M. H. Burkhardt, M. A. Hossain, S. Sarkar, Y.-D. Chuang, A. G. Cruz Gonzalez, A. Doran, A. Scholl, A. T. Young, N. Tahir, Y. J. Choi, S.-W. Cheong, H. A. Dürr, and J. Stöhr, Phys. Rev. Lett. **108**, 237202 (2012).

20. A. Yakubovskii, K. Kumagai, Y. Furukawa, N. Babushkina, A. Taldenkov, A. Kaul, and O. Gorbenko, Phys. Rev. B **62**, 9 (2000).

21. H. J. Lee, K. H. Kim, M. W. Kim, T. W. Noh, B. G. Kim, T. Y. Koo, S. W. Cheong, Y. J. Wang, and X. Wei, Phys. Rev. B **65**, 115118 (2002).

22 . J. L. Garcìa-Muńoz, A. Collado, M. A. G. Aranda, and C. Ritter, Phys. Rev. B **84**, 024425 (2011).

23. V. Siruguri, P. D. Babu, S. D. Kaushik, A. Biswas, S. K. Sarkar, M. Krishnan, and P. Chaddah, J. Phys.: Condens. Matter **25,** 496011 (2013).

24. A. Shahee, S. Sharma, K. Singh, N. P. Lalla, and P. Chaddah, AIP Conference Proceedings **1665**, 060004 (2015),

25 . Y. Imry, and M. Wortis, Phys. Rev. B **19**, 3580 (1979).

26 . A. Banerjee, K. Kumar, and P. Chaddah, J. Phys.: Condens. Matter **21**, 026002 (2009).

27 . F. Parisi, and L. Ghivelder, Physica B **398**, 184 (2007).